\documentclass[prb,twocolumn,preprintnumbers,amsmath,amssymb,floatfix,showpacs]{revtex4}

\usepackage[dvips]{graphics,graphicx,color}

\begin{document}

\title{Exciton-exciton interaction and biexciton formation in bilayer
systems}

\author{R.\ M.\ Lee, N.\ D.\ Drummond, and R.\ J.\ Needs}

\affiliation{TCM Group, Cavendish Laboratory, University of Cambridge,
J.\ J.\ Thomson Avenue, Cambridge CB3 0HE, United Kingdom}

\begin{abstract}
We report quantum Monte Carlo calculations of biexciton binding
energies in ideal two-dimensional bilayer systems with isotropic
electron and hole masses. We have also calculated exciton-exciton
interaction potentials, and pair distribution functions for electrons
and holes in bound biexcitons.  Comparing our data with results
obtained in a recent study using a model exciton-exciton potential
[C.\ Schindler and R.\ Zimmermann, Phys.\ Rev.\ B \textbf{78}, 045313
(2008)], we find a somewhat larger range of layer separations at which
biexcitons are stable.  We find that individual excitons retain their
identity in bound biexcitons for large layer separations.
\end{abstract}

\pacs{71.35.Cc, 78.67.De, 02.70.Ss}

\maketitle

\section{Introduction \label{sec:introduction}}

Electrons and holes in semiconductors can combine to form
hydrogen-like bound states called \textit{excitons}.  The creation and
recombination of excitons is one of the principal mechanisms by which
light interacts with semiconductors. Furthermore, excitonic systems
possess a number of unusual properties, such as the ability to
transport energy without transporting charge, suggesting a range of
applications in novel electronic devices. Excitonic systems have
therefore been the subject of numerous
experimental\cite{snoke_2002,butov_2002,butov_2004,rapaport_2004,hammack_2006}
and theoretical\cite{szymanska_2003,tan_2005,schindler_2008} studies
in recent years.

In the low-density limit, excitons may be regarded as
weakly-interacting neutral bosons.\cite{butov_2004} Bose-Einstein
condensation of excitons is therefore possible. However, it has proved
to be difficult to obtain exciton lifetimes that are sufficiently long
for thermalization to take place, and the current experimental
evidence for Bose-Einstein condensation is
inconclusive.\cite{butov_2003,snoke_2003} One promising approach for
overcoming the problem of short lifetimes is the development of
coupled-quantum-well (bilayer) systems, in which thin layers of
semiconductor and an applied electric field in the growth direction
are used to confine the electrons and holes to spatially separated,
parallel, quasi-two-dimensional wells, hindering recombination and
extending exciton lifetimes.\cite{butov_2002,snoke_2002,hammack_2006}

At present our understanding of the exciton-exciton interaction in
bilayer systems is limited.  On the one hand there is a repulsive
electrostatic interaction between excitons.  For example, if the layer
separation is nonzero then the excitons have parallel dipole moments,
giving an asymptotically dominant repulsive interaction.  Furthermore,
the static charge distribution of each exciton has a permanent
quadrupole moment in general (even at zero layer separation, provided
the electron and hole masses differ), giving another repulsive
interaction term.\cite{schindler_2008} On the other hand, fluctuating
dipole (van der Waals) forces result in an attraction between excitons
at short range.  Because of the existence of the van der Waals forces,
it is sometimes possible for biexcitons (bound states of pairs of
excitons) to form.  A better understanding of the interaction between
excitons in coupled quantum wells will facilitate the interpretation
of experimental data, in particular enabling the determination of the
exciton densities achieved in experiments.

The dependence of exciton and biexciton binding energies on the layer
separation has been investigated by Tan \textit{et al.}, who found
that, while the exciton binding energy decays slowly as the inverse of
the layer separation, the biexciton binding energy decays extremely
rapidly.\cite{tan_2005} Recent studies of the exciton-exciton
interaction using a heavy-hole approximation have found there to be a
critical layer separation for each electron/hole mass ratio, beyond
which biexcitons become unstable with respect to dissociation into two
separate excitons.\cite{zimmermann_2007,schindler_2008} In this
article we report quantum Monte Carlo (QMC) calculations of the
binding energies of biexcitons in bilayer systems and exciton-exciton
interaction potentials.

The rest of this article is arranged as follows.  In Sec.\
\ref{sec:biex_binding} we describe our calculations of the binding
energies of biexcitons and investigate the range of layer separations
and mass ratios for which biexcitons are stable.  In Sec.\
\ref{sec:xx_interaction} we present our data for the exciton-exciton
interaction potential. In Sec.\ \ref{sec:biex_pcf} we report
pair-distribution functions (PDF's) for biexcitons. Finally, we draw
our conclusions in Sec.\ \ref{sec:conclusions}.  We use Hartree atomic
units ($\hbar=|e|=m_e=4\pi\epsilon_0=1$) throughout this article,
although we report final energies in exciton Rydbergs ($R_y^\ast=\mu
e^4/[2(4\pi \epsilon_0 \epsilon)^2 \hbar^2]$, where $\mu=m_e m_h /
(m_e+m_h)$ is the reduced mass of an exciton and $m_e$ and $m_h$ are
the electron and hole masses) and lengths in terms of exciton Bohr
radii [$a_B^\ast= 4\pi \epsilon_0 \epsilon \hbar^2/(\mu e^2)$].

\section{Biexciton binding energies \label{sec:biex_binding}}

We have modeled the coupled-quantum-well system by an idealized
two-dimensional (2D) bilayer, in which the electrons and holes are
confined to two parallel planes, and the effective mass tensors of the
electrons and holes are isotropic. In reality, electrons and holes are
free to move within quantum wells that are of finite width (e.g., one
experimental setup\cite{butov_2004} has well widths of 8 nm and a well
separation of 4 nm), although the Coulomb attraction between electrons
and holes should keep the particles confined to the inner edges of
their respective wells. We have also restricted our attention to
biexciton systems in which the two electrons have opposite spins, as
do the two holes, because this is the ground-state spin configuration.

We have studied biexcitons and exciton-exciton interactions using the
variational and diffusion quantum Monte Carlo (VMC and DMC)
methods. In the VMC method the expectation value of the Hamiltonian
with respect to a trial wave function is calculated using a stochastic
integration technique.\cite{foulkes_2001} Trial wave functions usually
contain a number of free parameters, to be optimized by minimizing
either the energy expectation value or the variance of the energy. DMC
is a stochastic projector technique for solving the many-body
Schr\"odinger equation.\cite{ceperley_1980,foulkes_2001} DMC is in
principle exact for systems with nodeless ground-state wave functions,
such as the biexciton systems studied in this work.

Our trial wave function was similar to that of Tan \textit{et
al.},\cite{tan_2005} with additional flexibility provided by
multiplication by a two-body Jastrow factor.\cite{ndd_jastrow} This
wave function satisfies the Kato cusp conditions when particles
coincide,\cite{kato_pack} and reduces to the form appropriate for two
isolated excitons when the excitons are far apart.  We also carried
out some calculations using a three-body Jastrow factor.\cite{casino}
For a typical case where $E_{\rm VMC}-E_{\rm DMC}=3 \times 10^{-4}
R_y^*$, the reduction in the VMC energy from the inclusion of a
three-body term was $10^{-4} R_y^*$. Obtaining the best possible trial
wave function was especially important for the PDF calculations
described in Sec.\ \ref{sec:biex_pcf}.

We optimized the free parameters in our wave function by unreweighted
variance minimization\cite{umrigar_1988a,kent_1999,ndd_newopt} and
linear-least-squares energy minimization.\cite{umrigar_emin} The trial
wave function can describe the dissociated system more accurately than
it can describe the bound system; hence energy minimization is the
more sensible choice for investigating binding, although this depends
upon initial parameters and configurations. Although the DMC energy is
independent of the trial wave function, the statistical efficiency of
the method is increased when the wave function is improved. By using a
more flexible wave function, we have been able to achieve considerably
smaller error bars than Tan \textit{et al.}\cite{tan_2005} All the QMC
calculations reported in this article were performed using the
\textsc{casino} program.\cite{casino}

For each layer separation $d$ and electron/hole mass ratio $\sigma$,
the biexciton binding energy was calculated as $E_b=2E_X-E_{XX}$,
where $E_X$ is the energy of a single exciton and $E_{XX}$ is the
energy of the four-body biexciton system.  The exciton energy $E_X$
was obtained using a numerically exact Runge-Kutta integration
technique as described in Ref.\ \onlinecite{tan_2005}, while DMC was
used to calculate the biexciton energy $E_{XX}$. The DMC energies were
converged with respect to time step and population size; any remaining
bias is much smaller than the statistical error bars.

\begin{figure}
\begin{center}
\includegraphics[scale=0.3]{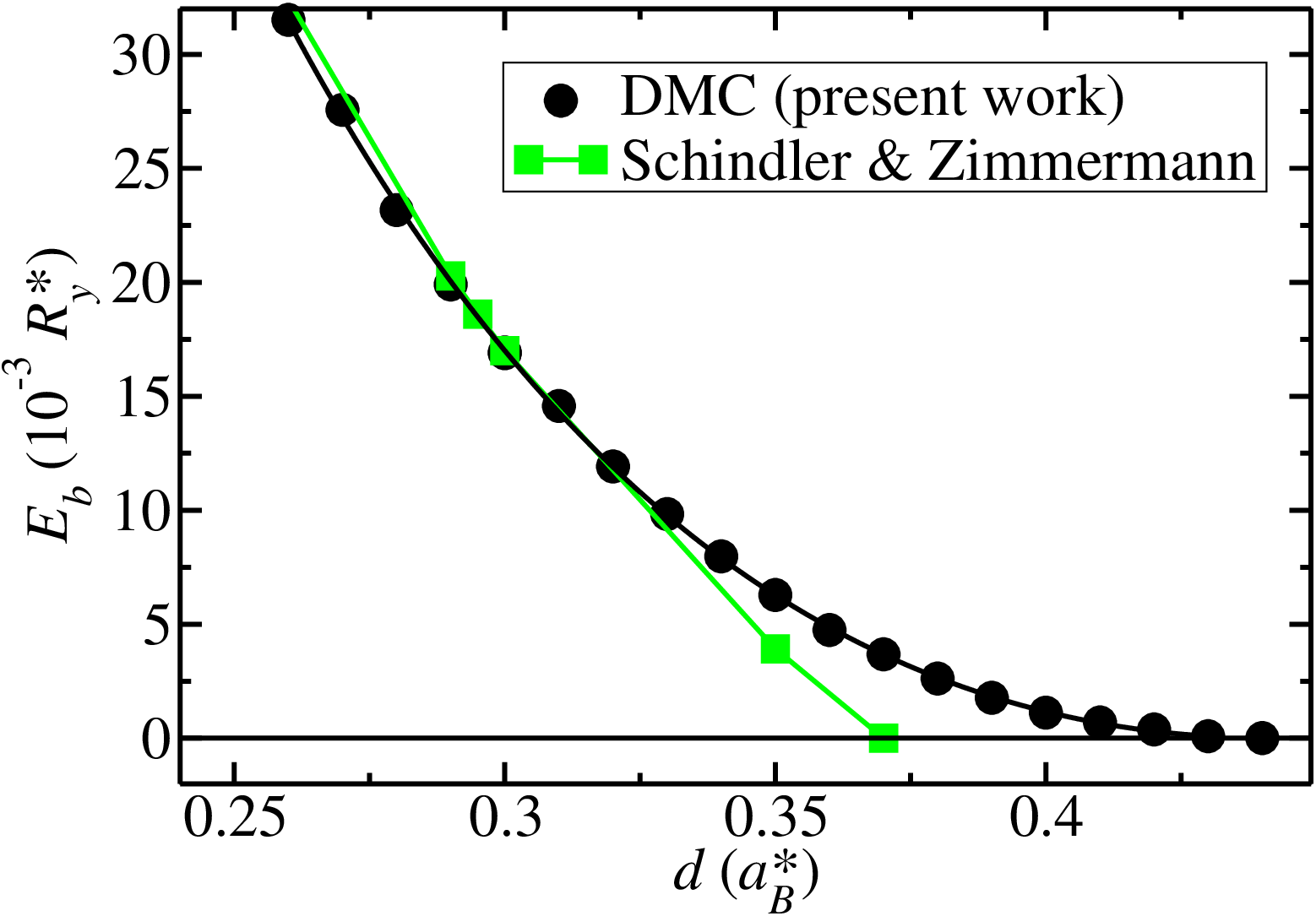} \\
\includegraphics[scale=0.3]{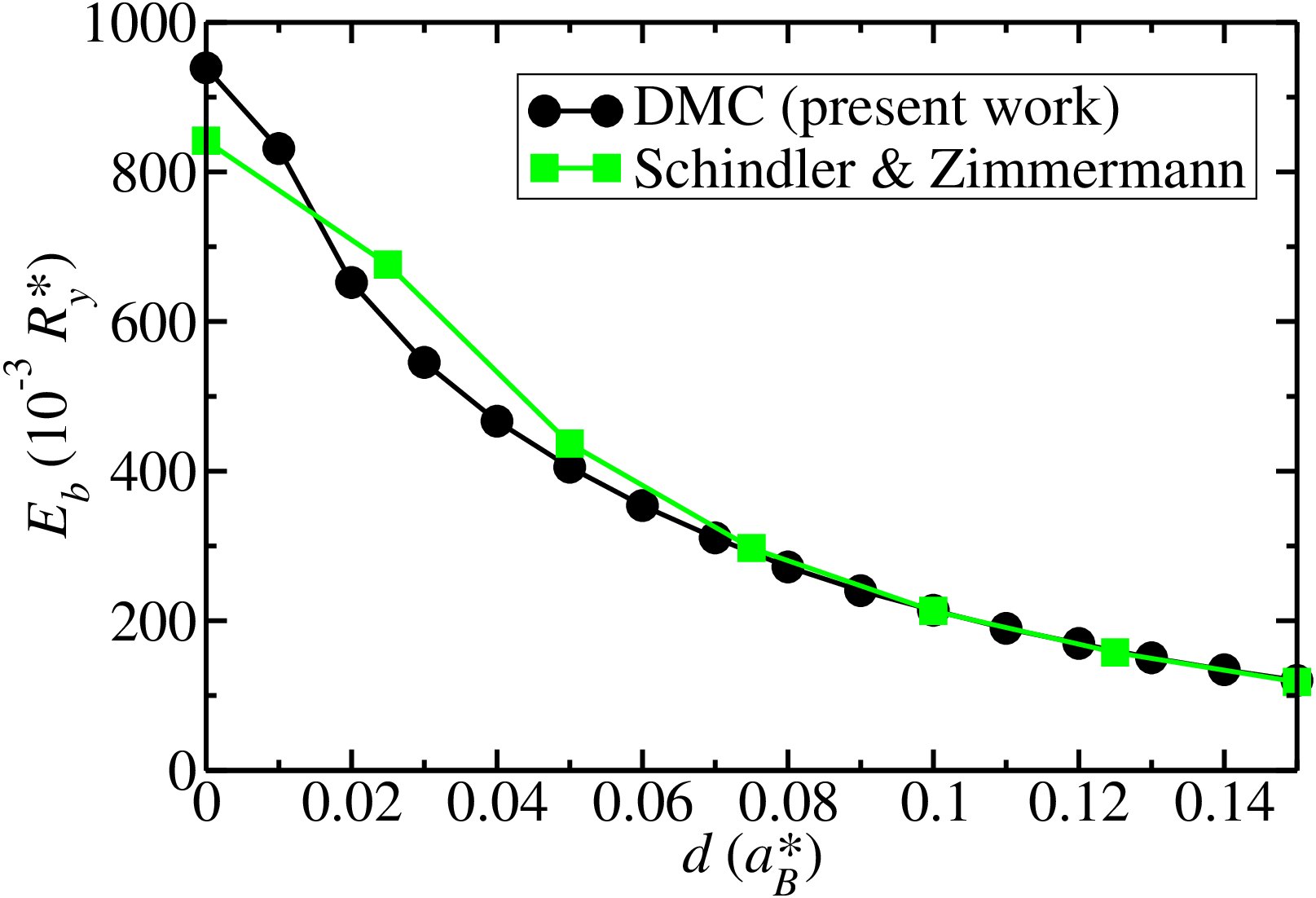}
\caption{(Color online) Biexciton binding energy $E_b$ as a function
of layer separation $d$ for electron/hole mass ratio $\sigma=0.3$.
The upper panel shows the binding energy for layer separations close
to the critical separation; the lower panel shows the binding energy
for small layer separations.
\label{fig:m03binding}}
\end{center}
\end{figure}

\begin{figure}
\begin{center}
\includegraphics[scale=0.3]{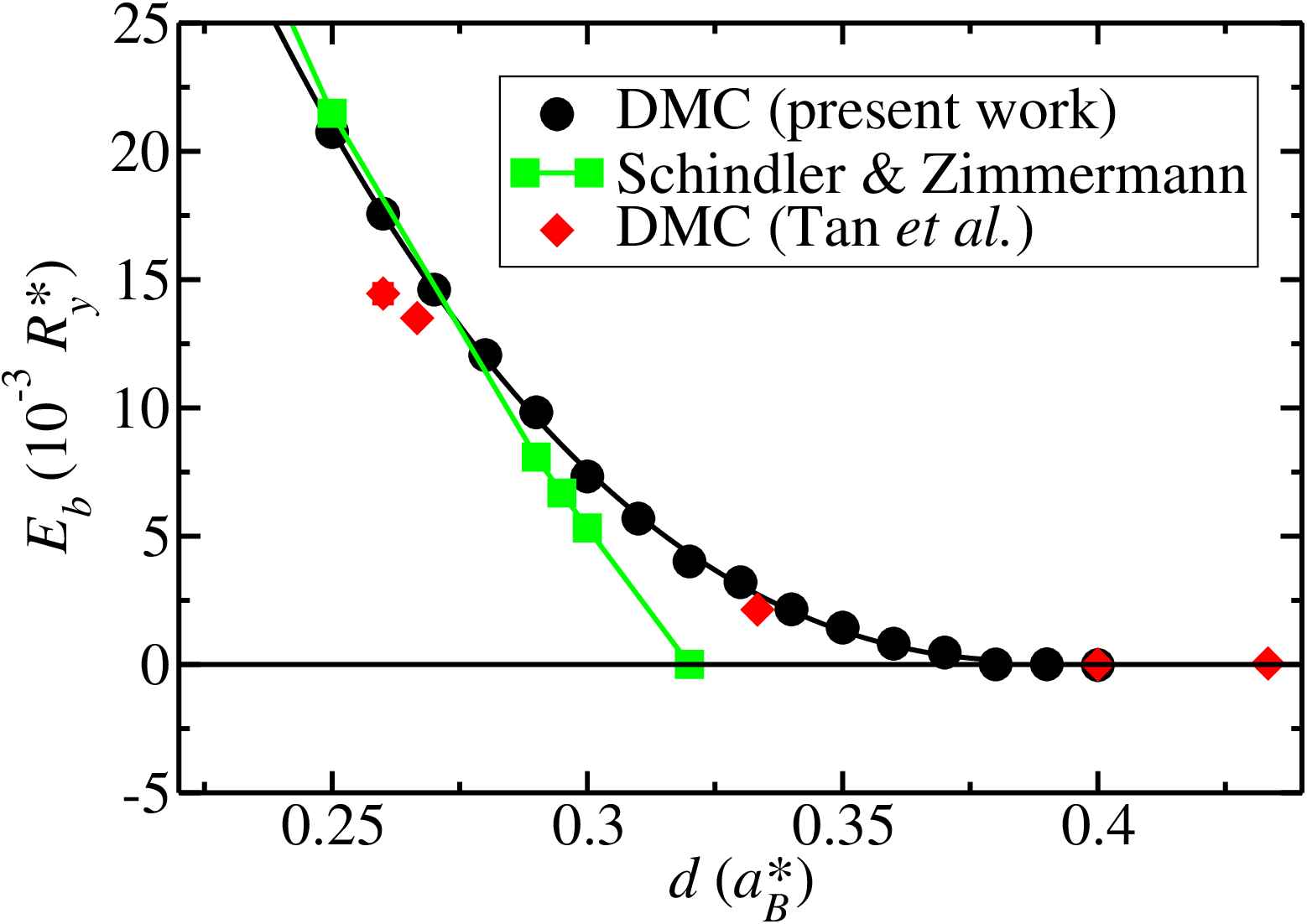} \\
\includegraphics[scale=0.3]{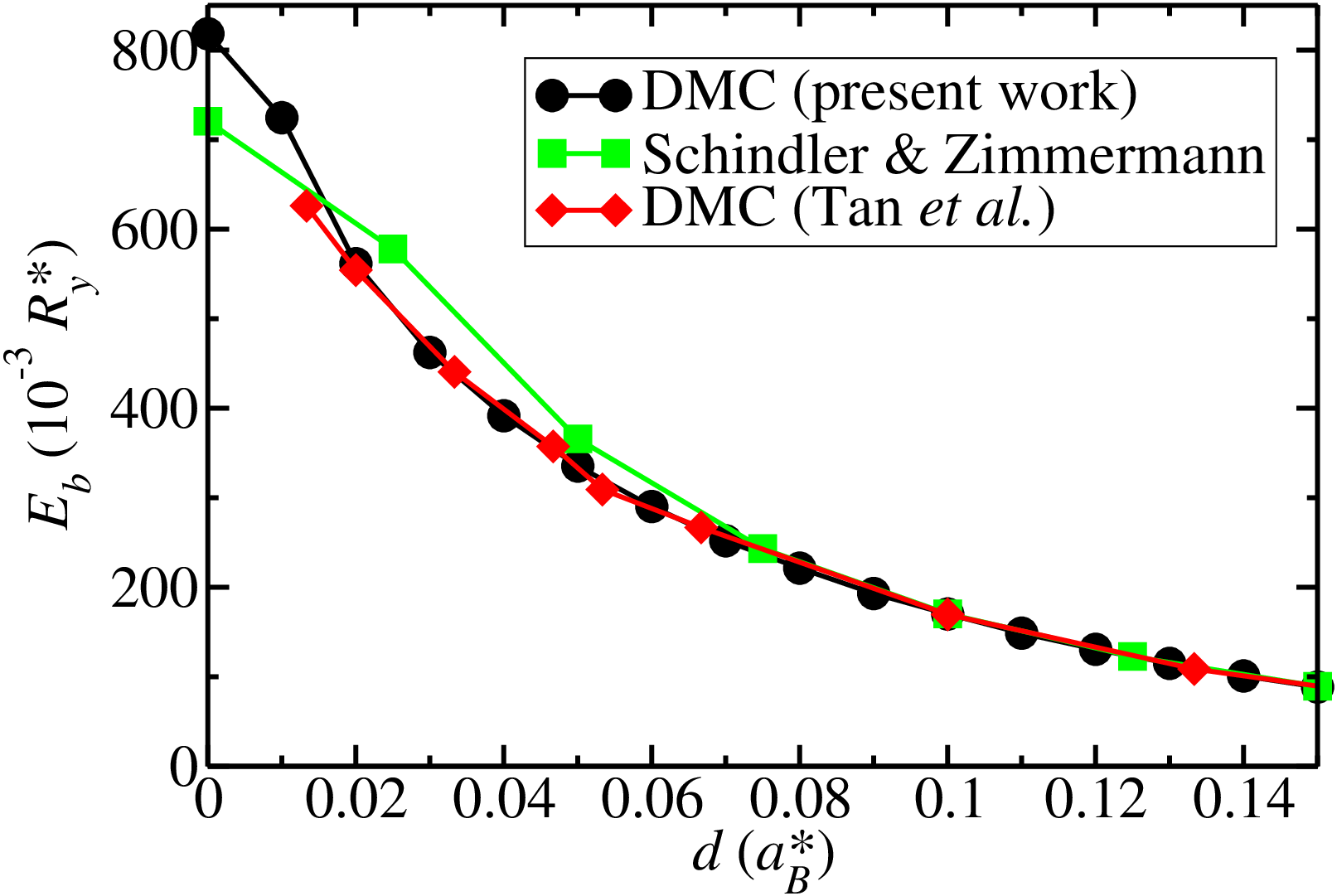}
\caption{(Color online) Biexciton binding energy $E_b$ as a function
of layer separation $d$ for electron/hole mass ratio $\sigma=0.5$.
The upper panel shows the binding energy for layer separations close
to the critical separation; the lower panel shows the binding energy
for small layer separations.
\label{fig:m05binding}}
\end{center}
\end{figure}

\begin{figure}
\begin{center}
\includegraphics[scale=0.3]{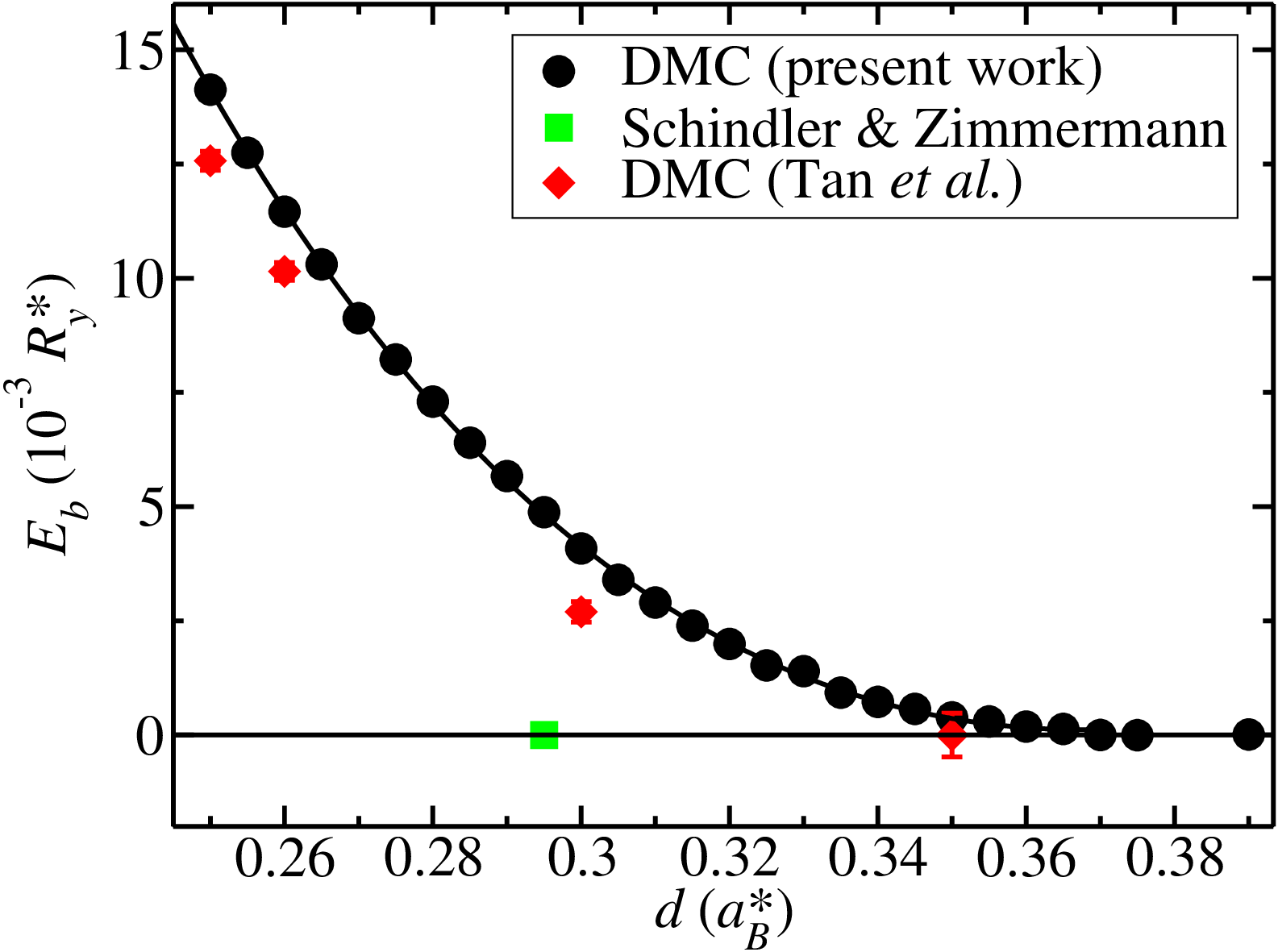}
\caption{(Color online) Biexciton binding energy $E_b$ as a function
of layer separation $d$ for equal electron and hole masses
($\sigma=1$).  The square shows Schindler and Zimmermann's estimate of
the critical point at which the biexciton ceases to be
bound.\cite{schindler_2008}
\label{fig:m10binding}}
\end{center}
\end{figure}

Biexciton binding energies for $\sigma=0.3$, $0.5$, and $1$ are shown
in Figs.\ \ref{fig:m03binding}, \ref{fig:m05binding}, and
\ref{fig:m10binding}, respectively. It can be seen that our results
are close to those of Tan \textit{et al.};\cite{tan_2005} the
difference arises from our use of exact single-exciton energies.  Tan
\textit{et al.}\ used Eq.\ (3) of Ref.\ \onlinecite{tan_2005} to
generate $E_X$ values, introducing a small, systematic error. Removing
this error reveals that our DMC data are in statistical agreement with
those of Tan \textit{et al}. The random errors in our data are much
smaller, so we can locate the layer separation at which the biexciton
ceases to be bound.  Tan \textit{et al.}\ fitted an exponential form
to their binding energy data, which resulted in the erroneous
conclusion that biexciton binding persists to infinite layer
separation.

Our DMC results show some deviation from the binding energies obtained
by Schindler and Zimmermann,\cite{schindler_2008} especially when
$d\rightarrow 0$, when $E_b\rightarrow 0$, and when the mass ratio is
close to 1, because  we have performed a full simulation of all four
particles in the biexciton,  whereas they simulated a pair of excitons
interacting via a model potential. The deviation of our binding
energies from those of Schindler and Zimmermann is approximately $4
\times 10^{-3} R_y^\ast$ where $E_b\rightarrow 0$. At smaller $d$ the
agreement is much better, but below $d\approx 0.1a_B^\ast$ we find
larger differences, reaching a maximum of almost $0.1 R_y^\ast$ at
$d=0$, as shown in Figs.\ \ref{fig:m03binding} and
\ref{fig:m05binding}.

\begin{figure}
\begin{center}
\includegraphics[scale=0.3]{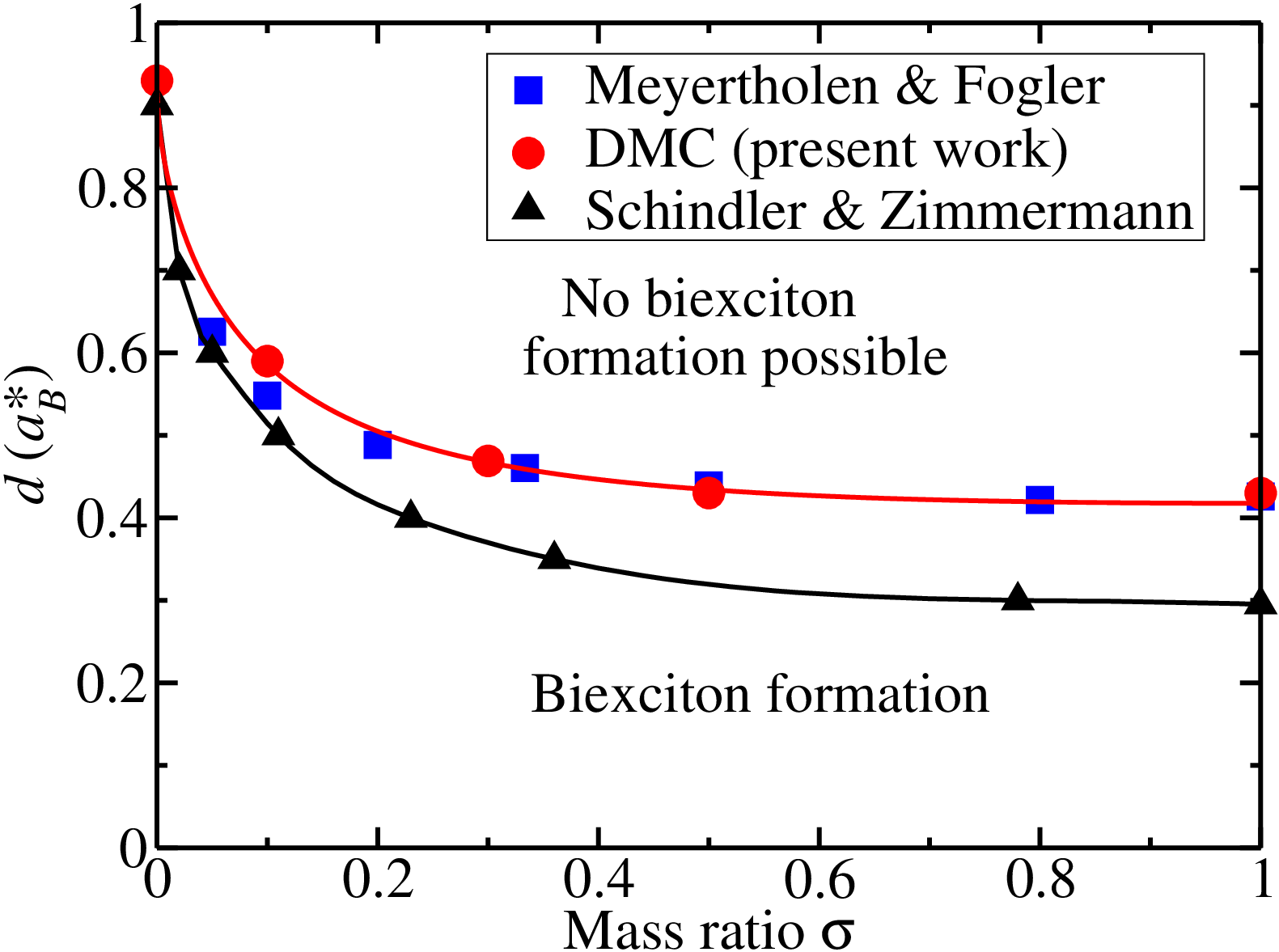}
\caption{(Color online) The region of biexciton stability from DMC
calculations compared with that found by Schindler and
Zimmermann\cite{schindler_2008} and Meyertholen and
Fogler.\cite{Meyertholen} The critical points were found by
extrapolating the biexciton binding energies to zero using the fitting
form set out in Ref.\ \onlinecite{Meyertholen}. The statistical errors
are comparable to the size of the symbols.
\label{fig:formation}}
\end{center}
\end{figure}

As can be seen in Fig.\ \ref{fig:formation}, which shows the range of
$\sigma$ and $d$ over which the biexciton is stable, we find a
somewhat  larger region of stability for the biexciton than Schindler
and Zimmermann.  Let $d_{\rm crit}(\sigma)$ be the critical layer
separation, beyond which the biexciton is unbound.  As
$\sigma\rightarrow 0$, the heavy-hole approximation made by Schindler
and Zimmermann becomes increasingly accurate, and our results for
$d_{\rm crit}(0)$ agree with theirs.  On the other hand, for
$\sigma=1$ their interaction potential is less accurate and our value
of $d_{\rm crit}(1)$ is therefore significantly higher than theirs.

Our data are mostly in excellent agreement with those of Meyertholen
and Fogler, although at small $\sigma$ we find a slightly larger
region of biexciton stability. This is not an artifact of the
extrapolation, which followed the scheme set out in Ref.\
\onlinecite{Meyertholen}, for we were able to find points with nonzero
binding energies outside the region of stability defined by
Meyertholen and Fogler. This is consistent with the variational
principle that applies to their results.

One may parameterize the boundary of the region of biexciton stability
in Fig.\ \ref{fig:formation}. Expressing $d_{\rm crit}$ in terms of
$\sigma+\sigma^{-1}$ ensures that the correct behavior is observed
upon exchanging the electron and hole masses [i.e., $d_{\rm
crit}(\sigma^{-1})=d_{\rm crit}(\sigma)$]. A suitable fitting function
is
\begin{equation}
d_{\rm crit}(\sigma)=\frac{F}{\sqrt{\sigma+\sigma^{-1}}}\, \tanh
\left[G\sqrt{\sigma+\sigma^{-1}}\right] + 0.93 \;,
\label{eqn:formationfit}
\end{equation}
where the parameter values $F=1.19(5)$ and $G=-0.50(4)$ give a
$\chi^2$ error of $0.4$ per data point. The functional form of Eq.\
(\ref{eqn:formationfit}) satisfies most of the conditions derived in
Ref.\ \onlinecite{Meyertholen}: $d_{\rm crit}^\prime(0)$ is infinite,
$d_{\rm crit}^\prime(1)=0$ and $d_{\rm crit}(0)-d_{\rm
crit}(\sigma)\propto \sqrt{\sigma}$ for $\sigma\ll1$.

\section{Exciton-exciton interaction \label{sec:xx_interaction}}

The exciton-exciton interaction potential $E_I(R)$ at separation $R$
is defined to be the energy of a biexciton system in which the centers
of mass of the two excitons are constrained to be a distance $R$
apart, minus the energies of two isolated excitons.  The Hamiltonian
for the constrained biexciton system may be written as
\begin{eqnarray}
\hat{H} & = & \frac{1}{2\mu}\left(\nabla_1^2+\nabla_2^2\right )
-\frac{1}{r_1}-\frac{1}{r_2} \nonumber \\ & & {} +
\frac{1}{|\mathbf{R}+\frac{\mu}{m_e} (-\mathbf{r}_2+\mathbf{r}_1)|} +
\frac{1}{|\mathbf{R}+\frac{\mu}{m_h}(-\mathbf{r}_1 +\mathbf{r}_2)|}
\nonumber \\ & & {} - \frac{1}{|\mathbf{R}-\frac{\mu}{m_h}
  \mathbf{r}_1-\frac{\mu}{m_e}\mathbf{r}_2|} -
\frac{1}{|\mathbf{R}+\frac{\mu}{m_e} \mathbf{r}_1+\frac{\mu}{m_h}
  \mathbf{r}_2|} , \nonumber \\
\label{eqn:com_H}
\end{eqnarray}
where ${\bf r}_1$ and ${\bf r}_2$ are the electron-hole separations
within the two excitons. The first two potential terms represent the
intra-exciton electron-hole potentials, followed by the hole-hole,
electron-electron, and finally the two inter-exciton electron-hole
terms.  DMC calculations can then be performed for an effective
two-particle system, with coordinates ${\bf r}_1$ and ${\bf r}_2$.
The kinetic-energy operator only includes derivatives with respect to
in-plane coordinates.  The form of trial wave function was the same as
that used in Sec.\ \ref{sec:biex_binding}, but with the electron and
hole coordinates being re-expressed in terms of ${\bf r}_1$, ${\bf
r}_2$, and the fixed vector ${\bf R}$.

\begin{figure}
\begin{center}
\includegraphics[scale=0.3]{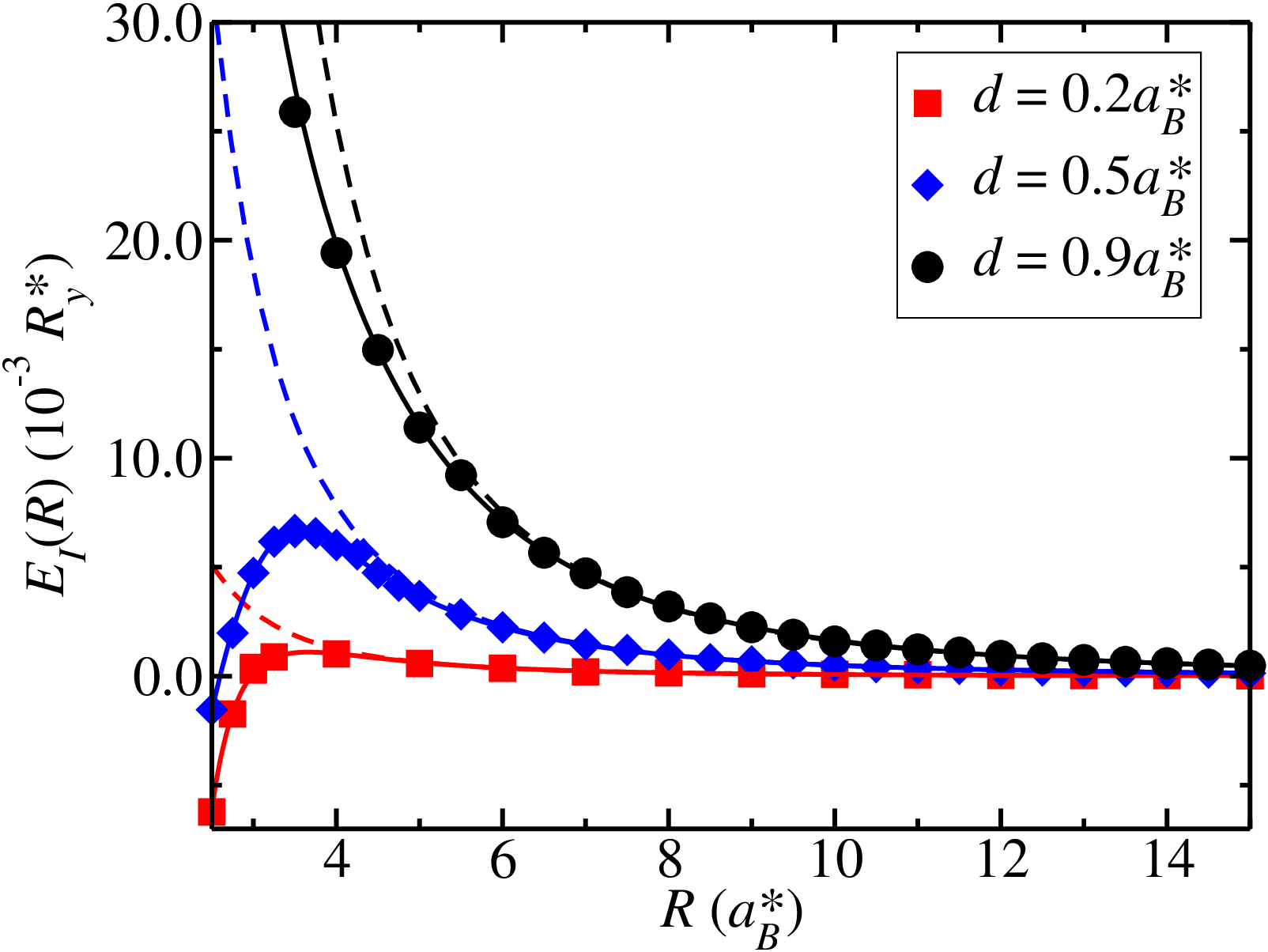}
\caption{(Color online) Exciton-exciton interaction potential $E_I(R)$
as a function of center-of-mass separation $R$ with $\sigma=1$. The
solid lines show the fit to the DMC data.  Dashed lines show the
dipole-dipole interaction energy [Eq.\ (\ref{eq:dip-dip})].
\label{fig:interaction_com}}
\end{center}
\end{figure}

The center-of-mass constraint may not be used to calculate the
interaction potential at zero exciton-exciton separation, because in
that limit the repulsion becomes strong enough to dissociate the two
individual excitons. The ground state of Eq.\ (\ref{eqn:com_H}) at
very small $R$ is thus not the physical quantity we require. We have
calculated the exciton-exciton potential only at separations $R$ for
which the excitons remain bound. Figure \ref{fig:interaction_com}
demonstrates this effect, exhibiting a potential which decreases at
small $R$ to physically unreasonable values.

Our DMC calculations yield a smooth exciton-exciton potential. The
interaction energies shown in Fig.\ \ref{fig:interaction_com} do not
deviate from simple dipole-dipole repulsion of the form
\begin{equation}
E_I(R)=\frac{2d^2}{R^3}\;,
\label{eq:dip-dip}
\end{equation}
by more than $3.5 \times 10^{-5} R_y^\ast$ above an exciton-exciton
separation of $\approx 7 a_B^\ast$. The fits to the  interaction
potential data are shown in Sec.\ \ref{sec:appendix}. The  repulsive
tails of the interaction ($R > 10a_B^\ast$) calculated for  pairs of
excitons with $0.1 < \sigma < 1$ all collapse onto a single  curve for
each value of $d$ when scaled into excitonic units, showing  a maximum
deviation from each other and Eq.\ (\ref{eq:dip-dip}) of  $8 \times
10^{-5} R_y^\ast$.

\begin{figure}
\begin{center}
\includegraphics[scale=0.3]{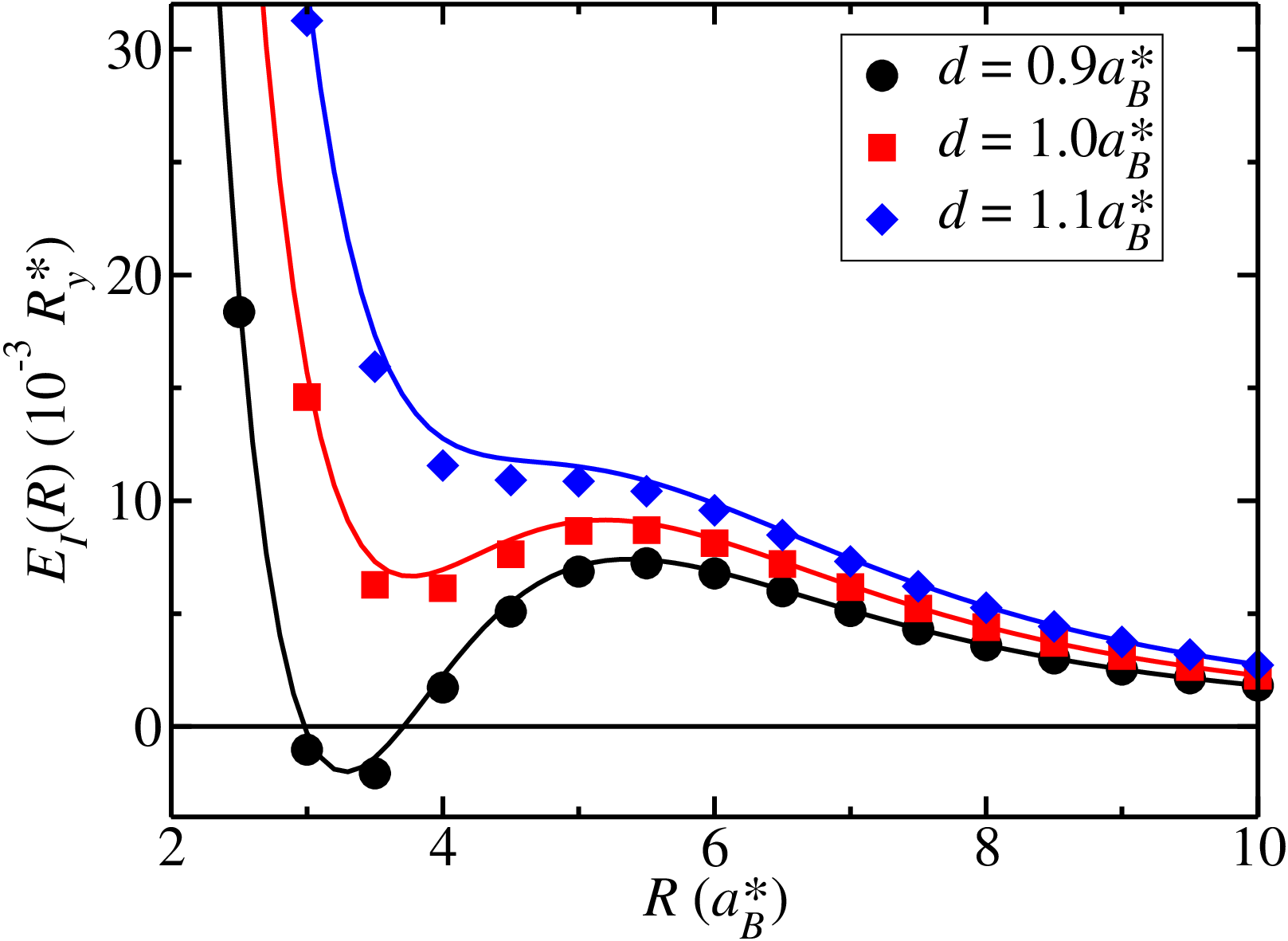}
\caption{(Color online) Exciton-exciton interaction potential $E_I(R)$
as a function of hole-hole separation $R$ for the heavy-hole case
($\sigma=0$) with $d=0.9$, $1.0$, and $1.1a_B^\ast$. The solid lines
show the interaction potential from Ref.\ \onlinecite{schindler_2008}.
\label{fig:interaction_holes}}
\end{center}
\end{figure}

For an electron/hole mass ratio of $\sigma=0$, our results should
reduce to the exciton-exciton interaction under the heavy-hole
approximation.\cite{schindler_2008} Figure \ref{fig:interaction_holes}
demonstrates the agreement with the interaction potentials calculated
by Schindler and Zimmermann. Our points are slightly below the curves
of Ref.\ \onlinecite{schindler_2008} at the potential minimum,
although the agreement is in general very good, and well within
statistical error at large $R$. Equation (\ref{eq:h-hfit}) (in Sec.\
\ref{sec:appendix}) shows a functional form suitable for fitting to
our data.

\begin{figure}
\begin{center}
\includegraphics[scale=0.3]{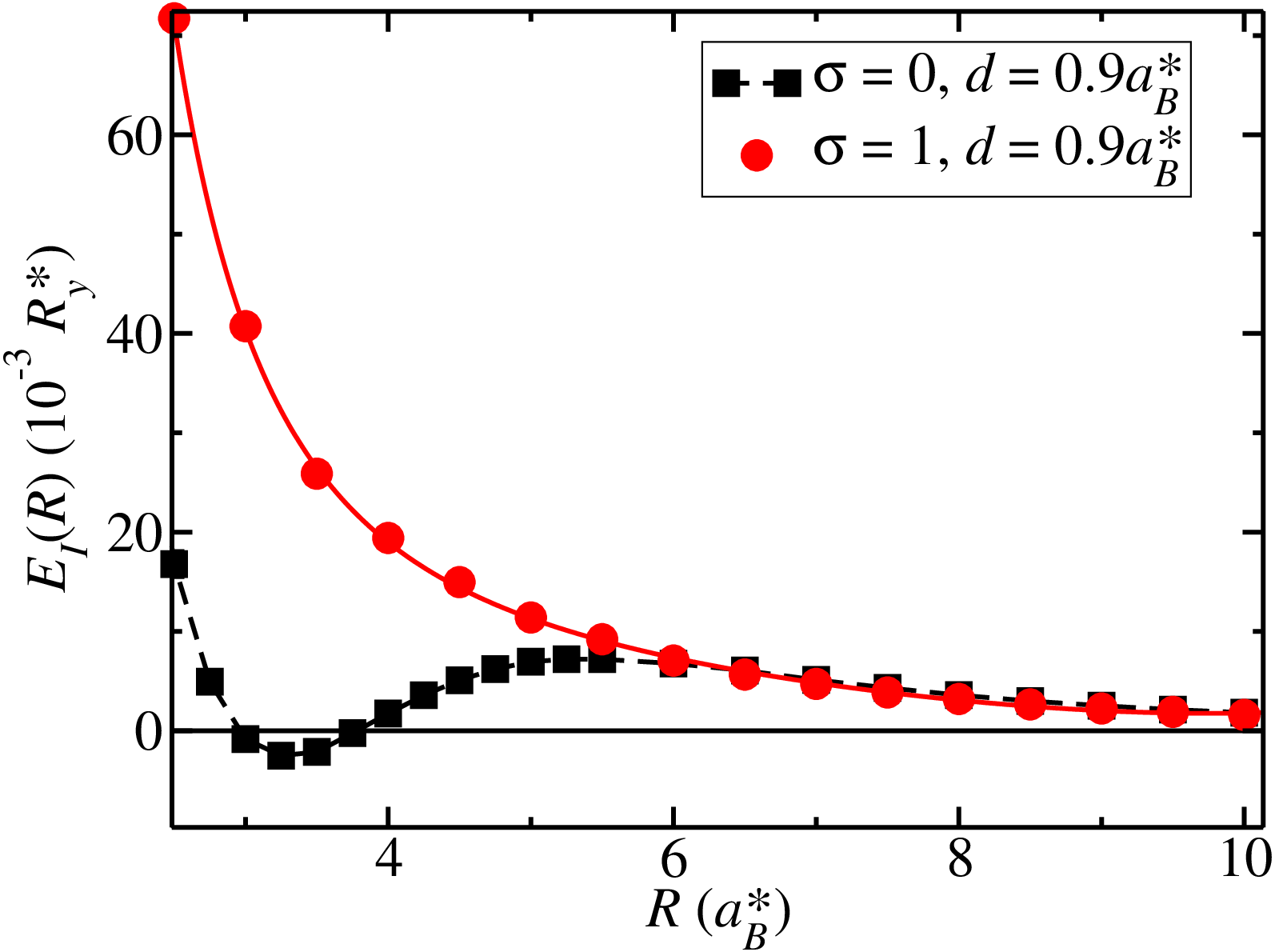}
\caption{(Color online) Exciton-exciton interaction potential $E_I(R)$
  as a function of (constrained) center-of-mass separation $R$ for
  $d=0.9a_B^*$ and $\sigma=0$ and 1.
\label{fig:interaction_both}}
\end{center}
\end{figure}

For large layer separations $d$ the interaction is purely repulsive,
whereas for smaller $d$ the interaction is attractive at short range.
The critical point in the binding occurs near the layer separation for
which the minimum in the exciton-exciton interaction potential
disappears. Schindler and Zimmermann's approach uses a model
exciton-exciton interaction potential which depends on the layer
separation $d$ but not the mass ratio $\sigma$.\cite{schindler_2008}
Constraining the center of mass rather than the hole positions allows
us to observe the interaction potential for different mass ratios and
layer separations, so that we do not need to apply the interaction
potential obtained in one system to another with different
parameters. The dependence of the interaction potential upon $\sigma$
is clear from Fig.\ \ref{fig:interaction_both}, and is consistent with
the results shown in Fig.\ \ref{fig:formation}, in which biexcitons
are stable at $d=0.9a_B^\ast$ for $\sigma=0$ but not $\sigma=1$.

For the strictly two-dimensional case ($d=0$), we can compare our
values  of the Haynes factor, $f_H=E_b/E_X$, with those of previous
work.  Usukura \textit{et al.}\ performed numerically exact
variational  calculations, finding $f_H=0.665$ for $\sigma=0$ and
$f_H=0.193$ for  $\sigma=1$.\cite{Usukura} These data agree well with
our values of  $f_H=0.670(3)$ and 0.19287(2) for $\sigma=0$ and 1,
respectively.

\section{PDF's in biexcitons \label{sec:biex_pcf}}

The PDF's of electrons and holes in biexcitons reveal important
information about the physics of biexciton binding.  The
electron-electron PDF is defined as
\begin{equation}
g_{\rm ee}(r) = \frac{1}{2\pi r} \left< \delta(|{\bf r}_{{\rm
    e}\uparrow}-{\bf r}_{{\rm e}\downarrow}|-r) \right>,
    \end{equation} where ${\bf r}_{{\rm e}\uparrow}$ and ${\bf
    r}_{{\rm e}\downarrow}$ are the positions of the up- and down-spin
    electrons and the angled brackets denote the average over sets of
    electron and hole coordinates distributed as the square of the
    ground-state wave function.  The hole-hole PDF is defined in a
    similar fashion.  The electron-hole PDF is defined to be
\begin{equation}
g_{\rm eh}(r) = \frac{1}{8\pi r} \left< \sum_{\sigma_{\rm
e},\sigma_{\rm h} \in \{ \uparrow,\downarrow \}} \delta(|{\bf r}_{{\rm
e}\sigma_{\rm e}}^\parallel-{\bf r}_{{\rm h}\sigma_{\rm
h}}^\parallel|-r) \right>, \end{equation} where ${\bf r}_{{\rm
e}\sigma_{\rm e}}^\parallel-{\bf r}_{{\rm h}\sigma_{\rm h}}^\parallel$
is the in-plane separation of an electron and a hole.  The PDF's may
be accumulated within QMC by binning interparticle distances.  The
errors in the VMC and DMC estimates of the PDF [$g^{\rm VMC}(r)$ and
$g^{\rm DMC}(r)$] are linear in the error in the trial wave function;
however, the error in the extrapolated estimate $g^{\rm
ext}(r)=2g^{\rm DMC}(r)-g^{\rm VMC}(r)$ is second order in the error
in the wave function.\cite{foulkes_2001} Our VMC and DMC PDF's are
very close to one another, so the errors in our extrapolated estimates
are small.  The PDF's presented here have been normalized such that
\begin{equation}
\int^\infty_0 2\pi r g^{\rm ext}(r) \, dr = 1.
\end{equation}
\begin{figure}
\begin{center}
\includegraphics[scale=0.3]{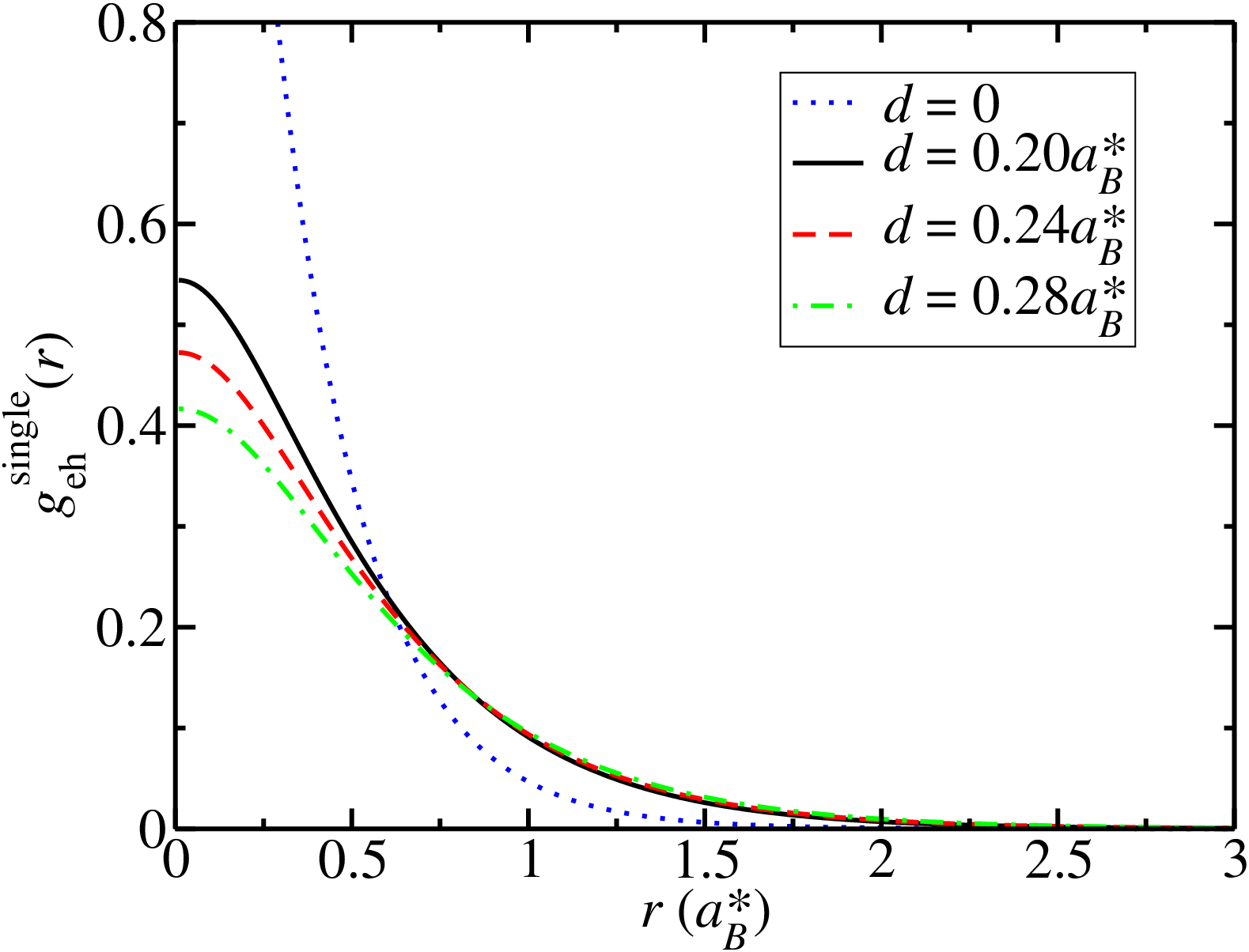}
\caption{(Color online) PDF $g^{\rm single}_{\rm eh}(r)$ for an
    isolated electron-hole pair from the exact solution of Eq.\ (2) in
    Ref.\ \onlinecite{tan_2005}, shown at several layer separations
    for  $\sigma=1$.
\label{fig:single_expdf}}
\end{center}
\end{figure}

\begin{figure}
\begin{center}
\includegraphics[scale=0.3]{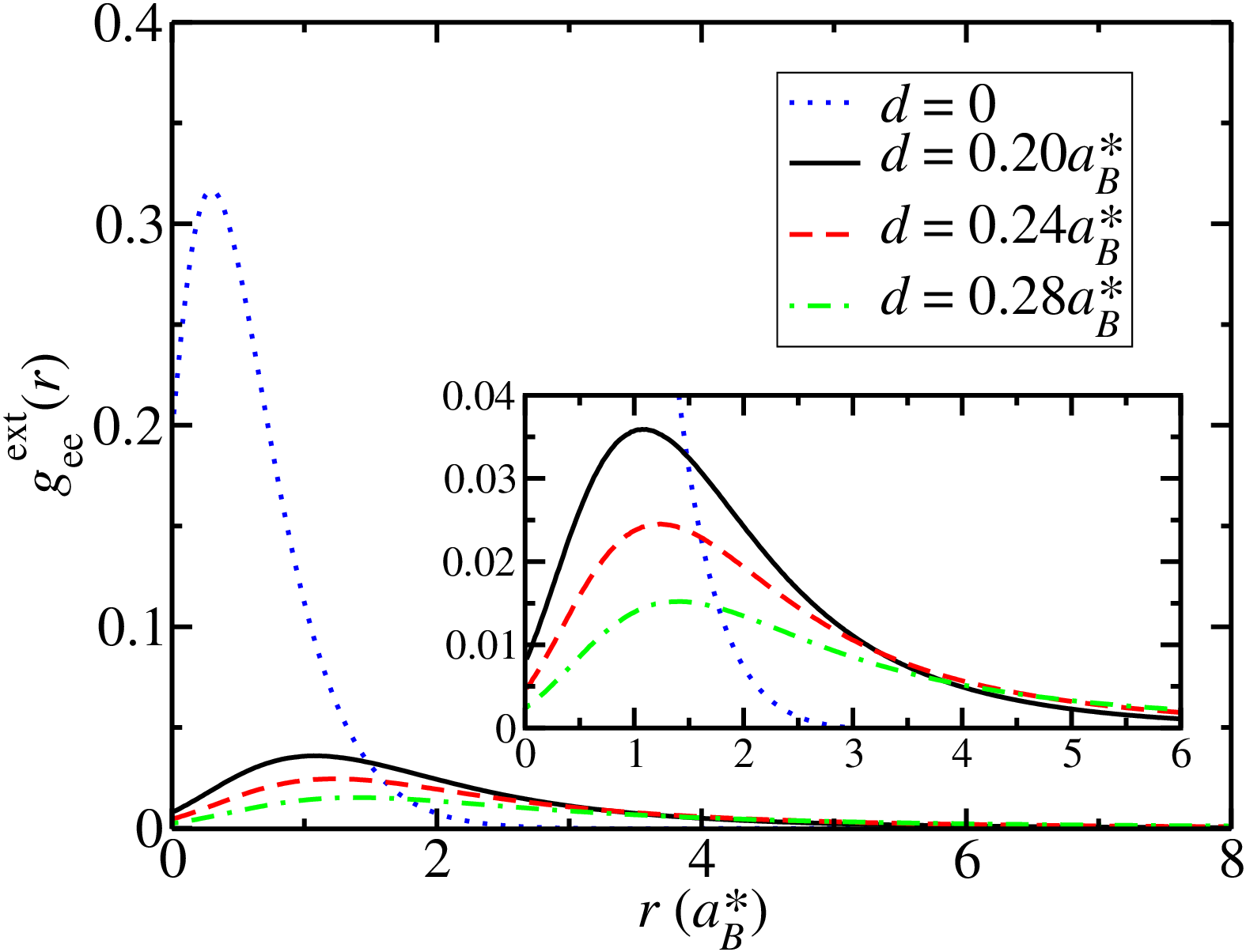}
\caption{(Color online) Extrapolated electron-electron PDF $g^{\rm
    ext}_{\rm ee}(r)$ for bound biexcitons with $\sigma=1$. The
    hole-hole and electron-electron PDF's are identical for equal
    electron and hole masses.
\label{fig:eepdfm10}}
\end{center}
\end{figure}

\begin{figure}
\begin{center}
\includegraphics[scale=0.3]{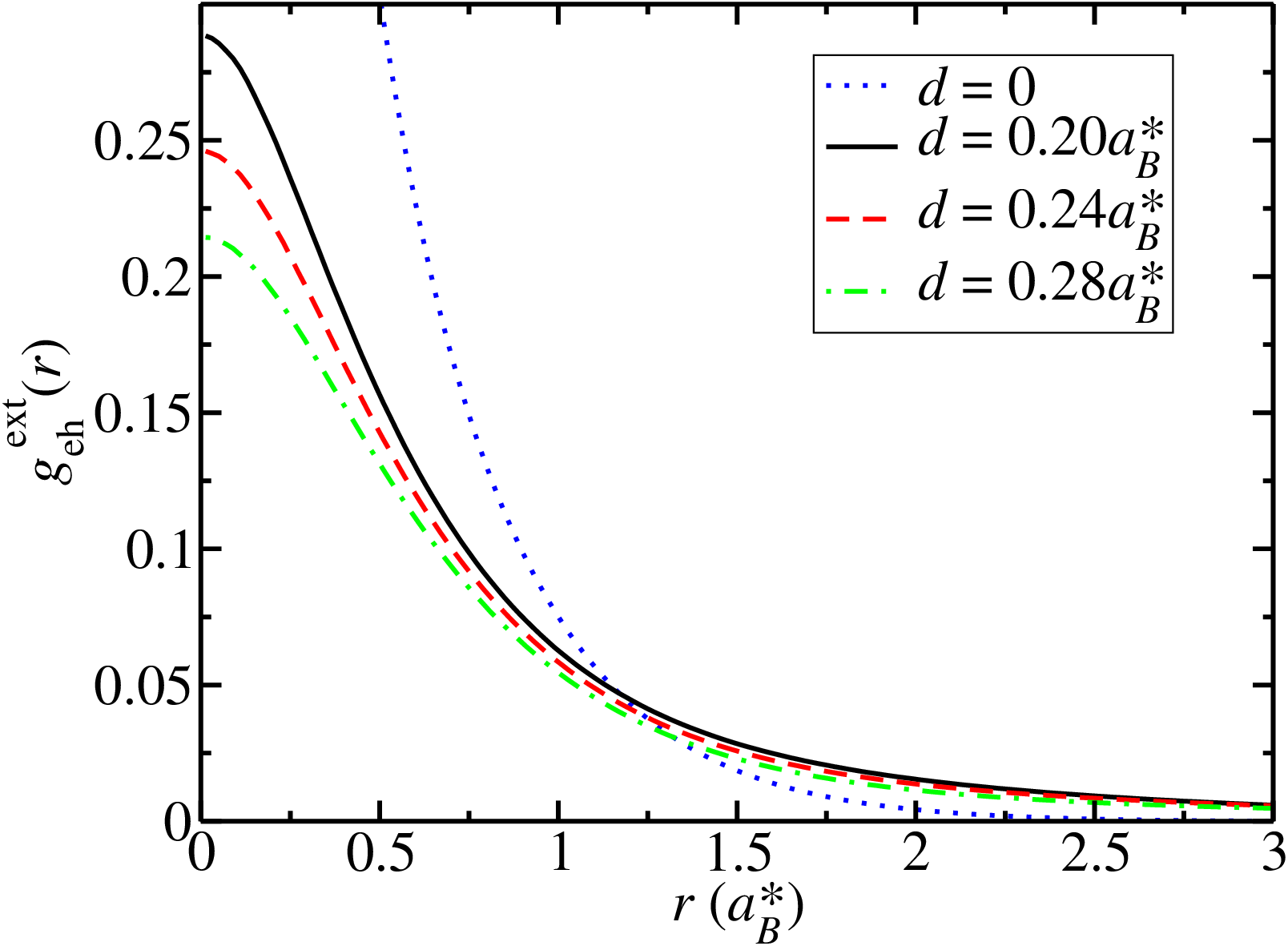}
\caption{(Color online) Extrapolated electron-hole PDF $g^{\rm
    ext}_{\rm eh}(r)$ for the biexciton system with $\sigma=1$ and
    several bilayer separations $d$.
\label{fig:ehpdfm10}}
\end{center}
\end{figure}

\begin{figure}
\begin{center}
\includegraphics[scale=0.3]{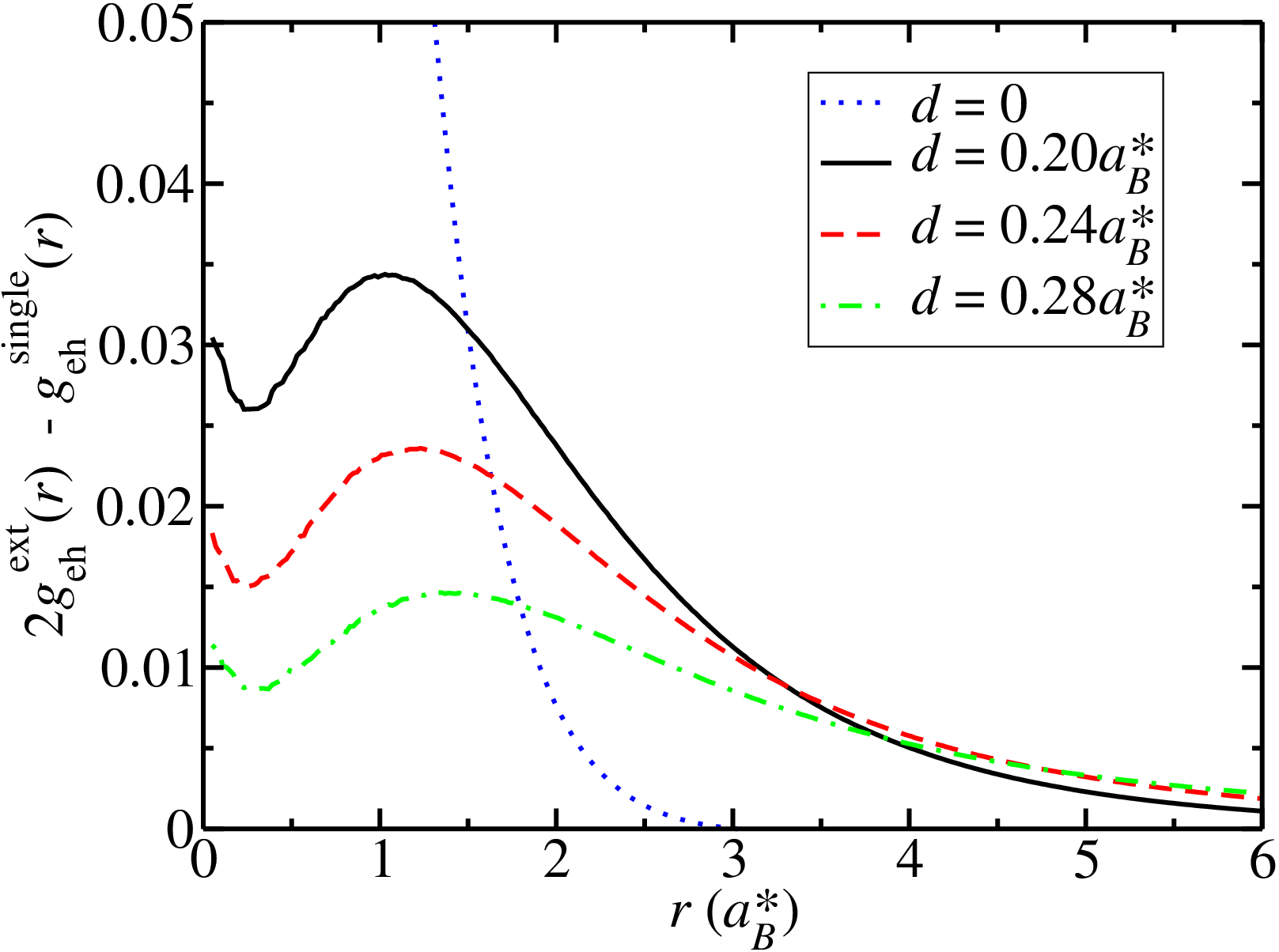}
\caption{(Color online) Biexciton electron-hole PDF relative to the
    single-exciton PDF, $2g^{\rm ext}_{\rm eh}(r)-g^{\rm single}_{\rm
    eh}(r)$, at $\sigma=1$ and several bilayer separations $d$.
\label{fig:pdf_diff}}
\end{center}
\end{figure}
Figure \ref{fig:single_expdf} shows the electron-hole PDF for a single
exciton, $g^{\rm single}_{\rm eh}$, obtained from the exact numerical
solution to Eq.\ (2) in Ref.\ \onlinecite{tan_2005}. Figures
\ref{fig:eepdfm10} and \ref{fig:ehpdfm10} show electron-electron and
electron-hole PDF's, respectively, for the biexciton system with
$\sigma=1$. At smaller layer separations the electron-hole PDF
exhibits a larger peak at zero interparticle separation, and decays
more rapidly with interparticle distance.

The size of the biexciton is most easily judged by examining the
electron-electron PDF (which is identical to the hole-hole PDF for
$\sigma=1$). The size of the biexciton diverges as the critical layer
separation ($d_{\rm crit}=0.43 a_B^\ast$ for $\sigma=1$) is
approached.  At zero layer separation, the electron-electron PDF is
negligible for interparticle distances larger than $3a_B^\ast$ and has
a maximum at $0.3a_B^\ast$.

Although a second peak cannot be discerned in Fig.\
\ref{fig:ehpdfm10}, the quantity $2g^{\rm ext}_{\rm eh}(r)-g^{\rm
single}_{\rm eh}(r)$ plotted in Fig.\ \ref{fig:pdf_diff} allows one to
see the inter-exciton electron-hole PDF superimposed on the change in
the intra-exciton PDF due to the presence of the other exciton. The
peaks in Fig.\ \ref{fig:pdf_diff} occur at the same separation as
those in Fig.\ \ref{fig:eepdfm10}, confirming that excitons retain
their identity in bound biexcitons for large layer separations, even 
when electrons and holes have equal masses. For zero layer separation 
there is no discernable peak, however, and the function rises sharply 
to a maximum at zero interparticle separation. This may be due to the 
large change in the single-exciton PDF due to the presence of the other 
exciton swamping the inter-exciton electron-hole PDF\@. We are thus 
unable to conclude with certainty that excitons retain their identities 
in bound biexcitons throughout the region of biexciton stability.

\section{Conclusions \label{sec:conclusions}}

We have carried out a QMC study of the interaction between pairs of
excitons in bilayer systems.  We have calculated the exciton-exciton
interaction potential by constraining the center-of-mass separation,
which we believe gives a more accurate pair potential at short range
than the potential calculated by assuming the holes to be infinitely
heavy.\cite{schindler_2008} We find that for large layer separations, 
excitons retain their identity when they bind to form a biexciton, 
suggesting that treating excitons as individual particles is a reasonable 
approximation. However, by solving the Schr\"odinger equation for all 
four particles in a biexciton, we find that the range of layer separations 
and mass ratios over which biexcitons are stable is somewhat larger than  
the region of stability predicted using exciton-exciton pair potentials.

\section{Acknowledgments}

Financial support has been provided by the UK Engineering and Physical
Sciences Research Council and Jesus College, Cambridge. Computing
resources have been provided by the Cambridge High Performance
Computing Service. We would like to thank C.\ Schindler for providing
the data shown in Figs.\ \ref{fig:m03binding}, \ref{fig:m05binding},
\ref{fig:m10binding}, and \ref{fig:formation}. We thank P.\ L\'opez
R\'{\i}os for assistance with the calculations.

\appendix

\section{Fit to the exciton-exciton potential \label{sec:appendix}}

\begin{table}
\begin{center}
\begin{tabular}{cccc}
\hline \hline

Parameter & $d=0.9a_B^*$ & $d=1.0a_B^*$ & $d=1.1a_B^*$ \\

\hline

$p_1$     & $-70.18$     & $-66.25$     & $-61.89$     \\

$p_2$     & $-4243$      & $-4538$      & $-4804$      \\

$p_3$     & $4296$       & $8422$       & $12560$      \\

$p_4$     & $8.086$      & $7.319$      & $6.420$      \\

$p_5$     & $21520$      & $30740$      & $40610$      \\

$p_6$     & $-15100$     & $-54120$     & $-98510$     \\

$p_7$     & $0.1284$     & $0.1451$     & $0.2424$     \\

\hline \hline
\end{tabular}
\caption{Coefficients appearing in Eq.\ (\ref{eq:h-hfit}) allowing the
reproduction of fits to the points shown in Fig.\
\ref{fig:interaction_holes}. Performing the fits using data with
$R\geq 3a_B^\ast$ yields $\chi^2$ errors of 0.79, 1.1 and 1.4 per
data point for $d=0.9$, $1.0$, and $1.1a_B^\ast$, respectively.
\label{table:hole_coeffs}}
\end{center}
\end{table}
The exciton-exciton potential curves with $\sigma=0$ in Fig.\
\ref{fig:interaction_holes} may be fitted to a function of the form,
\begin{eqnarray}
E_I &=&\left ( p_1+\frac{1000}{R}+\frac{p_2}{R^2} +\frac{p_3}{R^4}
\right ) \exp \left( -\frac{p_4 R^3}{1000} \right) \nonumber \\ & + &
\left ( \frac{2d^2}{R^3}+\frac{p_5}{R^5}+\frac{p_6}{R^6}  \right
)\left[ 1-\exp \left( -p_7 R^3 \right) \right],
\label{eq:h-hfit}
\end{eqnarray}
where $d$ is the layer separation and $p_1, \ldots, p_7$ are the
fitting parameters.  The function has the correct long-range behavior,
$E_I \propto 2d^2/R^3$ for $R \rightarrow \infty$. The fitting
parameter values are shown in Table \ref{table:hole_coeffs}.

The interaction potentials in Fig.\ \ref{fig:interaction_com} with
$\sigma=1$ may be fitted to a function similar to Eq.\
(\ref{eq:h-hfit}). This  time the form is
\begin{equation}
E_I=\left ( \frac{2d^2}{R^3}+\frac{p_1}{R^5}+\frac{p_2}{R^6}\right )
\left [ 1-\exp \left( - \frac{p_3 R^{p_4}}{1000} \right) \right],
\label{eq:comfit}
\end{equation}
\begin{table}[h]
\begin{center}
\begin{tabular}{cccc}
\hline \hline

Parameter & $d=0.2a_B^*$ & $d=0.5a_B^*$ & $d=0.9a_B^*$ \\

\hline

$p_1$     & $2302$       & $1463$       & $-6797$      \\

$p_2$     & $-8947$      & $-12580$     & $73200$      \\

$p_3$     & $1316$       & $5.813$      & $24.94$      \\

$p_4$     & $0.1123$     & $4.703$      & $2.465$      \\

\hline \hline
\end{tabular}
\caption{Coefficients appearing in Eq.\ (\ref{eq:comfit}) allowing the
reproduction of fits to the DMC results in Fig.\
\ref{fig:interaction_com}. Performing the fits using data with  $R\geq
2.5a_B^\ast$ yields $\chi^2$ errors of 1.55, and 1.07 per  data point
for $d=0.2$ and $0.5a_B^\ast$, respectively. The $\chi^2$ error is
larger for $d=0.9a_B^\ast$, the purely repulsive curve, but the
maximum deviation from the data points is only $1.2 \times 10^{-3}
R_y^\ast$.
\label{table:com_coeffs}}
\end{center}
\end{table} \\
where the long range behavior is once again reproduced correctly and
each  of the terms in the first bracket has a physical interpretation.
The $1/R^5$ term may be associated with quadrupole-quadrupole
repulsion and the $1/R^6$ term with van der Waals attraction. The
signs of the fitting parameters are consistent with this
interpretation for $d=0.2$  and $0.5a_B^\ast$. The parameter values
are shown in  Table \ref{table:com_coeffs}.

\end{document}